# Electronic structure of UGe$_2$ at ambient pressure: comparison with X-ray photoemission spectra

M. Samsel-Czekała[1*], M. Werwiński[2], A. Szajek[2], G. Chełkowska[3], R. Troć[1]

[1] *W. Trzebiatowski Institute of Low Temperature and Structure Research, Polish Academy of Sciences, Okólna 2, 50-422 Wrocław, Poland*

[2] *Institute of Molecular Physics, Polish Academy of Sciences, M. Smoluchowskiego 17, 60-179 Poznań, Poland*

[3] *A. Chełkowski Institute of Physics, Silesian University, Uniwersytecka 4, 40-007 Katowice, Poland*

* Corresponding author e-mail: m.samsel@int.pan.wroc.pl; tel. +48 71 3954 322, fax. +48 71 344 10 29

Based on experimental crystallographic data, electronic structure of UGe$_2$ have been calculated and compared with our results of X-ray photoelectron spectroscopy (XPS) measurements. We employed two different advanced full potential (FP) methods: FP-local-orbital (FPLO) and FP-linear augmented plane waves (Wien2k) codes for non-magnetic and ferromagnetic states. Starting from the local spin-density approximation (LSDA) or generalised gradient approximation (GGA), we verified either the orbital polarisation (OP) correction or the GGA+$U$ approach for the U 5f-electrons, changing Coulomb-repulsion energies $U$ in the range 0 - 4 eV. Satisfying agreement was achieved between experimental and our calculated magnetic moments using *ab-initio* LSDA+OP and *non-ab-initio* GGA+$U$ approaches, the latter for realistic $U$ values of 2 - 3 eV. We proved by the LSDA+OP approach an existence of the Fermi surface nesting vector along the *a* axis, possibly responsible for the triplet superconducting pairing. The calculated data reveal predominantly an itinerant U 5f-electron character of bands near the Fermi level, E$_F$, with only small contributions from the U 6d and Ge 4p states. The experimental XPS spectrum of valence bands (VB) also contains the sharp main 5f-electron peak at E$_F$, a wide hump (around -2 eV), and broad small peaks at higher energies. In the calculated XPS spectrum, the width of the main 5f-electron peak varies between 0.8 and 1.4 eV, depending on a method used in computations, but the hump remains unresolved. A newly observed asymmetric 1-eV satellite in the experimental 4f-core XPS spectrum together with known 3-eV and 7-eV satellites suggest *dual* behaviour of U-5f-electrons in UGe$_2$, the feature is inferred also from the VB studies.

**Keywords:** A. magnetic intermetallics;  B. electronic structure of metals and alloys; B. magnetic properties; E. electronic structure, calculation

## 1. Introduction

UGe$_2$ by many authors is treated as having ferromagnetic (FM) order (T$_C$ = 52 K) due to *itinerant* 5f electrons which are believed to form also the superconducting (SC) state under the pressure region of (1.0 - 1.6) GPa [1]. The discovery of the latter has opened a number of speculations as to the formation of the unconventional SC-state in this compound coexisting with a fairly strong FM ordering. It appears that the SC-transition temperature, $T_s$, determined at about 1.2 GPa, is about 0.8 K. At the same time, the Curie temperature and ordered moment are 35 K and about 1 $\mu_B$, respectively [1, 2]. In such a situation, merely the *nonunitary triplet state* of the superconductivity is possible [3]. However, the physics of UGe$_2$ is still a matter of wide debates in the literature where researchers try to solve this unusual problem of the FM/SC coexistence discovered for the first time.

In this Introduction, we restrict ourselves mainly to underlining the electronic properties of UGe$_2$ at ambient pressure. Just at this pressure, the existence in UGe$_2$ of another characteristic temperature $T^* \sim \frac{1}{2} T_C$ was inferred from a wide maximum observed at the temperature derivative of the resistivity [4, 5] and a broad hump both in the specific heat [6, 7] and at thermal expansion [4, 7, 8]. The magnetic fluctuations, ascribed to this anomaly have been regarded until now as playing an essential role in the mechanism of superconductivity. A coupled charge and spin density wave (CDW/SDW) transition was suggested as the origin of the T*-anomaly [9]. However, no satellite peaks of the nesting origin were observed in neutron diffraction measurements [2,10]. Recently Kuwahara et al. [11] have found a very broad elastic anomaly related to $T^*$ in the longitudinal $c_{11}$ mode. Similarly a very broad huge negative minimum in the magnetoresistivity (MR) of the order of 40%, measured for the j||b and B||a configuration, which indicates the presence of very strong magnetic fluctuations around $T^*$, was also reported previously [12]. Therefore, now this is the reliable fact that we do not have to do with the second-order transition at $T^*$, as proposed in [9], at least at ambient and low pressure region. This anomaly presents rather some *cross-over* behaviour (see e.g. [5, 12]). It is also interesting to note that at temperatures close to $T^*$, a sudden delocalisation process of 5f electrons takes place, as inferred from the Hall-effect study [13]. All these results together with very recent and detailed investigation of single crystalline UGe$_2$ using the positive muon spin rotation technique [14] support the view that in UGe$_2$, within the temperature range of the FM state, two U 5f electron subsets (dualism) exist, differing much in their localisation character (see arguments provided in [5, 14]). Such a scenario has also been proposed by Yaresko et al. [15] in their considerations on the electronic structure of this unusual compound (see below). From the theoretical point of view, the idea that some U 5f states are localised while others remain itinerant had been reported much earlier by Miyake and Kuramoto [16].

Considerations, based on another mechanism, where all the U 5f electrons in UGe$_2$ are treated as having an itinerant character, are predominant in the literature (see e.g. [17] and [18]). Since the problem of the FM/SC coexistence in UGe$_2$ is one of the most important issues of the last decade, a reliable and detailed recognition of its electronic structure plays in this aspect a profound role. Unfortunately, despite having published till now more than one hundred papers, concerning a behaviour of this unusual uranium compound, there still exist many unclear problems which require an application of different modern approaches, now being available.

Among the large set of papers, also some number of them have been devoted to band structure calculations of UGe$_2$ [19-22], which will be discussed in Section 3.2. Previously, some experimental electronic structure investigations employing ultraviolet photoelectron spectroscopy (UPS), X-ray photoelectron spectroscopy (XPS) using Mg K$\alpha$ radiation (1254 eV), and bremsstrachlung isochromat spectroscopy (BIS) with an overall instrumental resolution with a poor average value of 0.5(2) eV, were reported for polycrystalline UGe$_2$ [23, 24]. Also polycrystalline ingots of UGe$_2$ were used in photoelectron measurements by the application of synchrotron radiation [25]. More recently the results of temperature dependent (within 14–100 K) UPS measurements with as high-resolution as 10 meV have been performed on single-crystalline UGe$_2$ [26]. The spectrum showed a sharp peak near the Fermi level (E$_F$) which gradually moved toward E$_F$ with increasing temperature in agreement with an expectation for itinerant U 5f-electrons. However, no temperature changes have been observed above 100 meV of binding energy (BE), which may reflect the presence in UGe$_2$ of more localised 5f-electrons at higher BEs.

In turn, Yaresko et al. [15] have reported X-ray absorption and magnetic circular dichroism (XMCD) measurements performed at the M$_{4,5}$ edges of uranium in UGe$_2$. The obtained spectra were also estimated by electronic structure calculations. They discussed several variants, taking into account the results of three groups of authors presenting in the literature: i) photoemissions [23-26] ii) electron-positron momentum densities (EPMD) [27] and iii) angular dependence of the de Haas-van Alphen (dHVA) frequencies [28-30].

The determined Fermi surface (FS) by the band structure calculations [20, 22, 27] and the dHvA effect in UGe$_2$ [28-30] is nearly cylindrical along the the *b* axis. The cyclotron masses m$_c$ were found to be relatively large (2 – 25 m$_0$), expected in the cases of a medium electronic specific heat coefficient $\gamma(0)$ (= 30 - 35 mJ K$^{-2}$ mol$^{-1}$) [2, 28-31]. The value of the coefficient of the electrical resistivity [5, 32] was typical of a Fermi liquid material (A~0.01 µΩcm/K$^2$). Based on the results of low-temperature MR, UGe$_2$ was found to be a compensated metal with the open orbits existing just along the *b* axis [28-30].

In this paper, we present results of various kinds of band structure calculations, being analysed in detail in Section 3.2, and compared to own results of XPS (Al K$\alpha$) measurements of single-crystalline UGe$_2$.

## 2. Experimental and computational details

Single crystals of UGe$_2$ have been grown by the Czochralski-pulling method in a tetra-arc furnace. Starting materials were uranium (99.98% wg) and germanium (99.999% wg). X-ray examination of the single crystal gave results as the ZrGa$_2$ type of the orthorhombic structure (*Cmmm*) and the lattice parameters, being consistent with earlier measurements [5, 33].

The XPS spectra in a broad range of 0 – 1400 eV BE were obtained with monochromatic Al K$_\alpha$ radiation 1486.6 eV at room temperature (RT) using a PHI 5700/660 Physical Electronic Spectrometer. The energy spectra of the photoelectrons were analysed by a hemispherical mirror analyser with an energy resolution equal to about 0.3 eV. The Fermi level was referred to the Au 4f$_{7/2}$ line for which the 4f-level BE is 84.0 eV. All spectra were measured immediately after breaking the sample in vacuum of $10^{-10}$ Torr. To avoid oxygen contamination in the experimental chamber, the samples were repeatedly cleaved *in situ* about every 10 min. We did not observe any increase of the oxidation effect during the data acquisition time. It was checked by observations of the (O 1s) spectra before and after each measurement. On this basis, we are convinced about rather negligible contamination of our sample by oxygen. The (C 1s) peak was also observed in the investigated spectra.

The electronic structure calculations have been performed within density functional theory (DFT) [34] using the full-potential local-orbital minimum-basis band structure scheme (FPLO) in its versions 5 and 7 [35] and the full-potential linearised augmented plane wave (FP-LAPW) method implemented in the latest version (Wien2k) of the original Wien code [36]. The FPLO calculations were performed in fully-relativistic mode and the LSDA exchange-correlation potential was assumed in the form proposed by Perdew and Wang [37]. It should be emphasized here that the fully relativistic version of the FPLO code, where the four-component Kohn-Sham-Dirac equation is solved, treats exactly all relativistic effects, including the spin-orbit (SO) interaction - without approximations. It is not the case for the Wien2k code where the scalar relativistic approach was implemented with the SO interaction taken into account approximately by employing the second variational method [38]. In the latter code, two different exchange-correlation potentials in the generalised gradient approximations (GGA) were tested as proposed by Perdew et al. [39] as well as Wu and Cohen [40].

Furthermore, to improve a description of the strongly correlated U 5f electrons by the Wien2k code, the on-site Coulomb repulsion energy *U* and exchange *J* parameters were introduced within the GGA+*U* approach [41] in the version introduced by Anisimov et al. [41(a)] with an approximate self-interaction correction (SIC) implemented in the rotationally invariant way according to Liechtenstein et al. [41(b)]. A detailed discussion on expression for total energy and the double counting term specification in GGA+*U* methods implemented in Wien2k code is given in [41(c)].

The most important *ab initio* approach, applied by us to minimise discrepancy between the calculated total magnetic moments, $\mu_{tot}$, in the LSDA approach and experimental ones, $\mu_{exp}$, and simultaneously predict the proper magnetization axis, is one that takes into account so-called orbital polarisation (OP) term. It was proposed by Brooks and Eriksson et al. [42-44] and implemented in both FPLO and Wien2k methods.

The number of ***k***-points was 8000 in the Brillouin zone (BZ), which corresponds to 1100 points in irreducible wedge of the BZ for all methods of calculations applied in this paper. For the BZ integration a tetrahedron method was used [45]. The self-consistent criterion was equal to at least $10^{-6}$

Ry for the total energy. In the case of the Wien2k code, the cut-off parameter used to determine the number of plane waves needed for the expansion of the wave function $R^{MT}_{min} \times K_{max} = 8$, where $R^{MT}_{min}$ is the smallest muffin-tin spherical radius present in the system and $K_{max}$ is the magnitude of the largest K vector. The calculations were performed for the lattice parameters and atomic positions in the unit cell (u.c.) as given in [33]. In all cases, DOSs and FSs were computed. Furthermore, to evaluate also the effect of magnetocrystalline anisotropy, the calculations were performed assuming magnetisation arranged along all main crystallographic axes. For these directions, values of magnetocrystalline anisotropy energy (MAE) were determined.

The simulated XPS spectra were computed based on the calculated partial DOSs for each atomic orbitals, multiplied by the proper photoelectron cross sections [46] and then convoluted by Gaussian of the full width at half maximum (FWHM), δ, equal to 0.3 eV that corresponds to the energy resolution of the experiment. This procedure neglects an evaluation of energy-dependent transition matrix elements as done, e.g., in [47]. As the simulated spectrum relies on tabulated [46] values of the photoemission cross sections, calculated up to 2 significant digits (for partial cross sections used in the present paper), the overall accuracy of the photoemission intensity is not poorer than a few percent, which is quite enough in the aspect of agreement with our experiment. However, only relative intensities of the features/peaks of the photoemission spectrum may be affected by uncertainties of the partial cross sections used in the calculations, not their positions on the energy scale.

## 3. Results and discussion

*3.1. Crystal and magnetic structures*

The ZrGa$_2$-type crystal structure (*Cmmm*) of UGe$_2$ consists of antiphase zigzag chains of U atoms lying in the *ab* plane and running along the [100] crystallographic direction. The nearest interchain and intrachain U–U distances are comparable to each other, amounting to about 0.39 nm. The structure, where the U central atom is coordinated by 8 closer lying and 2 more distant Ge atoms [5], possesses inversion symmetry. In the u.c., the U atom has one while Ge atoms as many as three non-equivalent crystallographic positions being as follows: U(4j), Ge1(4i), Ge2(2a), and Ge3(2c). Only U and Ge1 have their free parameter y, given at RT or 60 K, based on X-ray [33] or neutron diffraction [48] investigations, respectively. The neutron powder diffraction examination, performed below $T_C = 52$ K, yielded a collinear magnetic structure with the uranium ferromagnetically ordered moments arranged along the zigzag in the [100] direction [48, 33].

*3.2. Band structure and Fermi surface*

Band structure results obtained using two different approaches, FPLO [35] and Wien2k [36] codes, are presented. The difference in the results based on the Wien2k package for two different GGA exchange-correlation potentials, as proposed by Perdew et al. [39] and Wu and Cohen [40], turned out to be irrelevant.

At first, we employed the LDA and GGA approaches, which gave DOSs generally similar to one another. In Fig. 1 we show in detail the total DOS and its decomposition into atomic contributions (part (a)) and partial DOSs per electron orbitals (parts (b) and (c)) obtained by the FPLO code.

One can infer from Fig. 1 that the valence bands, occurring in the energy range from -11.5 to -5.3 eV, are formed predominantly by the Ge 4s electrons which hybridise with the U 6d electrons, yielding a pseudogap at about -8.3 eV. These valence bands are separated by a small gap of 0.4 eV width from other bands cutting $E_F$. Hence, the bottom of the latter conduction bands is situated at about -4.9 eV. In turn, the conduction bands are formed in their whole energy range by the hybridised U 5f with U 6d and Ge 4p electrons and they all together create a metallic bond.

Furthermore, as is visible in Fig. 1(b), the main contribution to the total DOS from -1.3 eV up to $E_F$ comes from the U $5f_{5/2}$ electrons (about 85-90%) in terms of a sharp peak separated by a SO pseudogap from the U $5f_{7/2}$ peak being centred at 1 eV above $E_F$. Successively, in the energy range from -4.9 eV to -1.3 eV, the total DOS consists of contributions from the strongly hybridised Ge 4p with U 6d electrons and in smaller degree also with the U 5f electrons. Interestingly, the Ge 4s valence electrons are separated from the Ge 4p conduction ones by a gap of about 0.8 eV width (see Fig. 1(c)).

Next, we applied the GGA+*U* approach, to simulate strong electron-electron correlation and improve agreement between simulated XPS spectrum and our experimental one (discussed in Sect. 3.3), implemented in the Wien2k package. Unexpectedly, all these calculations yielded minor changes with respect to the LDA results described above. All they in more illustrative are shown below in Fig. 7.

On the next step, to reproduce values of $\mu_{exp}$ in UGe$_2$, spin-polarised computations, utilising as above the FPLO and Wien2k codes, were performed within the LSDA or GGA schemes. The results differ slightly from each other, depending on the functional used in calculations. Nevertheless, even GGA produces too low values of $\mu_{tot}$ of about 0.3 $\mu_B$ on an uranium atom like does LSDA in these and previous calculations [15, 20-22] with respect to the observed ground state with $\mu_{exp}$ of about 1.5 $\mu_B$ [2]. Interestingly, all these calculations reproduced properly the easy magnetisation axis along the [100] direction by minimising the total energy. The calculated linear term, $\gamma_b$, of the electronic specific heat, i.e. Sommerfeld coefficient, is varying between about 15 and 17 mJ K$^{-2}$ mol$^{-1}$ for the spin polarisation along [100]. Experimental values $\gamma(0)$ are estimated between 30 - 35 mJ K$^{-2}$ mol$^{-1}$ [2, 28-31] and the relation $\gamma(0)/\gamma_b$ indicating a dynamic enhancement arising from e.g. magnetic fluctuations with possible contributions also from charge fluctuations is equal to 1.2 - 2.3. The above results revealed a well-known drawback of LSDA and also GGA approaches that an orbital magnetic moment is usually underestimated within these schemes. Thus, our next step was to introduce an additional orbital polarisation (OP) term within the LSDA+OP and GGA+OP functionals [42-44]. This option was possible in both cases of FPLO and Wien2k codes. The obtained values of $\mu_{tot}$, collected in Table 1, are now higher than those inferred from the LSDA or GGA approaches (without OP or $U$). For spin and orbital polarisation along the [100] direction, which is again indicated as the easy magnetisation axis (see MAE in Tables 1), the obtained moments are equal to 1.61 and 1.56 $\mu_B$/U atom for FPLO and Wien2k, respectively. Thus, the latter value is very close to $\mu_{exp}$ of 1.42 - 1.5 $\mu_B$/U atom, taken from [48, 33] (see Fig. 4). It should be emphasised that the LSDA+OP or GGA+OP methods used are *ab initio* and do not require any external parameters as Coulomb $U$ and exchange $J$ that have to be introduced within the LSDA+$U$ or GGA+$U$ approaches. Surprisingly, the calculations in the LSDA+OP scheme gave, however, much lower values of $\gamma_b$ than without the OP term. This is connected with the fact that for the LSDA+OP approach the Fermi level cuts DOS in a broad dip with strongly reduced intensities, located between two peaks [see Fig 2(a)]. Such a camel-shaped DOS was expected from model calculations just for the UGe$_2$ superconductor [17]. It was proposed that the T$^*$ anomaly, characterized e.g. in [5], may originate from such a camel-shaped DOS [17] or a Stoner gap $\Delta$ of a perfectly polarised state below T$^*$ [18]. It is worth underlining here that in all cases presented in Figs. 2 and 3, the DOSs at $E_F$ in spin-up channels have several times higher intensities than those in the spin-down ones, which makes UGe$_2$ being close to possessing the half-metallic property. The same feature was concluded from the LSDA+$U$ approach results presented previously in [20, 21].

As mentioned in the Introduction, the dHvA experiments suggested an itinerant but strongly correlated character of the U 5f-electron states because of high cyclotron masses of carriers [28-30]. Most of earlier band structure calculations, which used the LSDA+$U$ approach only, took into account just this character of the U 5f electrons. Both Yaresko et al. [15] and Shick et al. [20, 21] used the LSDA+$U$ approach but either in the linear muffin-tin orbitals (LMTO) in the atomic sphere approximation (ASA) or FP-LAPW methods, respectively. To get a satisfying agreement between $\mu_{tot}$ and $\mu_{exp}$, they had to use considerably small values of $U$ (= 0.5 and 0.7 eV) and the same $J$ (= 0.44 eV). However, these values of $U$ and $J$ differ from atomiclike ones, i.e. 2 eV and 0.55 eV. At first, we performed analogous calculations but employing the Wien2k code in the GGA+$U$ approach, taking the values of $U$ and $J$ parameters as those in [20]. As a result, our $\mu_{tot}$ was substantially lower (1.09 $\mu_B$/U atom) than $\mu_{exp}$ and $\mu_{tot}$, reported in [20] and [15], i.e. 1.46 and 1.57 $\mu_B$/U atom, respectively (see Table 1 in [15]). Hence, we had to increase values of the $U$ and $J$ parameters to reach better agreement with the experiments. The obtained results are collected in Table 2 and for better visualisation we have plotted them in Fig. 4 together with the results of works [15, 20]. The corresponding DOSs for $U$ = 2.0 eV are displayed in Figs. 2(b) and 3 (b,d). The calculations for $U$ = 1.5 eV and $J$ = 0.55 eV gave the same value of $\mu_{tot}$ (1.27 $\mu_B$/U atom) for both directions [100] and [010], but the latter one becomes now the easy magnetisation axis, which, of course, is inconsistent with experiments. In turn, by taking the atomiclike values (given above) in our calculations, again $\mu_{tot}$ is too small compared with $\mu_{exp}$. Only for the values: $U$ = 3 eV and $J$ = 0.55 eV, we have obtained $\mu_{tot}$ = 1.41 $\mu_B$/U, being equal to the measured neutron value [33] along the [100] direction. It turned out that $\mu_{exp}$, established by both magnetisation [5, 48, 28] and neutron diffraction [2, 10, 48, 33] measurements, is contained in a narrow range of values 1.4–1.5 $\mu_B$, marked by a hatched strip in Fig. 4. By inspecting this figure, it is clear that the curve for Wien2k (GGA+$U$) tends to a gradual saturation starting already from $U > 1$ eV and reaches the best agreement with the experiment for $U$ near a value of 3 eV. However, for $U > 3$ eV a rapid upturn deviation from this tendency has been observed. As is also seen, our curve up to $U = 3.0$ eV has a similar shape, compared to that reported in [15], but yields considerably smaller values of $\mu_{tot}$.

It is worth to emphasize that employing the GGA+$U$ functional, we obtained good agreement between $\mu_{tot}$ and $\mu_{exp}$ for realistic values of the $U$ parameter of 2-3 eV, contrary to that for LSDA+$U$ ($U = 0.6$ eV) given in [20]. The latter value of $U$ is certainly too small and has no justification. For instance, perfect agreement with the experimental moments has recently been reported by Yu et al.[49] for UO$_2$ just by using $U = 2$ eV in calculations [(see AFSO case in Fig. 4 in [49]]. These authors also used a wide range of $U$ values varying between 0 and 4.5 eV, to find the proper $U$ value, but in the LSDA+$U$ approach. Also several authors of earlier papers on various uranium compounds [44,50,51] applied successfully just $U = 2$ eV and $J = 0.5$ eV in their calculations.

The Fermi surface of the non-magnetic (LDA) state of UGe$_2$, obtained using FPLO code (not displayed here), exist in three Kramers, double-degenerated, conduction bands numbered 133, 135, and 137. In the 135th band, a large hole FS sheet occurs being open along the direction parallel to the $b$ axis, in agreement with the MR measurements [28-30]. However, contrary to the MR results, we do not observe any open orbits along the $a$ axis (in the C*mmm* space group).

As expected, both papers by Settai et al. [52] as well as Biasini and Troć [27] reported that, under the assumption that the 5f electrons are itinerant (as in this work), the non-magnetic FS of UGe$_2$, obtained by the LAPW method within LDA, exists in three bands and is in overall agreement with our results.

The FS sheets of UGe$_2$ in the ferromagnetically ordered state (FM-FS) along the lattice direction $a$ ([100]), obtained by us in the LSDA approach (not shown), originate from four non-degenerated bands (135, 136, 137, and 138). Interestingly, in the 136 band a large hole FS sheet exists having more pronounced quasi-two-dimensional (Q2D) character along the $b$ axis, in accordance with the MR data [28-30]. Next, in the middle 137 band, there is a large electron FS sheet containing an open structure along the $a$ axis, again in accord with the cited MR results. As to the unrealistic FM-FS of UGe$_2$ based on LSDA, so far only Settai et al. [52] and Yamagami [22] have reported their LAPW results - FS existing also in four bands. Unfortunately, we are not able to compare our data with the results given in [52]. The reason is that the FS sections are drawn in a completely different convention from ours. On the other hand, the FS sections, visualised in [22] with probable nesting vectors that could be connected with a formation of CDW/SDW, are different from our ones.

In order to obtain the FS in UGe$_2$ for realistic magnetic moments, we had to introduce the orbital polarisation correction in the LSDA+OP approach within the FPLO code. As a result, the FM-FS of UGe$_2$ along the [100] direction this time consists of three sheets coming from non-degenerate bands, 136, 137, and 138. In the latter band, there are solely very small electron pockets (not displayed here).

Fig. 5 shows the Q2D hole FS sheet in the 136 band, being open along the *b* axis. The sheet is almost the same as that in our LSDA case but accompanied by small hole pockets, cigars, and discs. Finally, in the 137 band (not displayed), there is a large electron FS sheet possessing a complex shape. The FS of UGe$_2$, calculated in the LDA+*U* approach in the work [20], exists also in three conduction bands. Their largest hole FS sheet (see Fig. 3 in [20]), having Q2D properties, is very similar to ours as drawn in Fig. 5(a), although it was obtained by a different approach. A section in *ac* plane of this Q2D-surface is displayed in Fig. 5(b) with a possible nesting vector **q** along the *a* axis of a length 0.47(2π/*a*), which is nearly equal to 0.45(2π/*a*) reported in [20]. Interestingly, these Q2D FS sheet (having a predominantly majority spin character) with nesting feature may be responsible for longitudinal magnetic fluctuations which, in turn, mediate in p-wave (triplet) superconducting pairing for equal spin states in UGe$_2$.

*3.3. XPS-experiment and theory*

In our XPS spectrum of Al Kα type, taken in the wide energy range of 0 – 1400 eV (not shown), only a small amount of oxygen and carbon was seen for BE of 531 eV (O 1s) and 283.5 eV (C 1s) respectively. In addition, the former small peak was obscured by some Ge LLM line. Thus, we conclude that the presence of uranium dioxide in our s.c. sample was certainly negligible.

The XPS spectrum up to BE of 33 eV is displayed in Fig. 6(a), where except for the valence band (VB) region, the positions of the SO split, $\Delta_{SO}$= 8.4 eV, U 6p$_{1/2}$ and U 6p$_{3/2}$ peaks as well as the Ge 3d peaks are also shown. The latter (centred at about 29 eV BE) has been deconvoluted and is also shown in the inset of Fig. 6(a) in the extended BE scale. As seen, there are two peaks which are split by $\Delta_{SO}$= 0.6 eV. It is interesting to note that each of the three Ge (1,2,3) atoms in the u.c. has the same contribution to our calculated XPS spectrum (not shown here) in a manner of four quite narrow peaks originating from the Ge 3d electrons that form two groups of very close lying doublets, separated by the above $\Delta_{SO}$. However, the average position of this complex Ge 3d peak structure is shifted by almost 5 eV towards lower BE compared with the experiment. Moreover, in Fig. 6(b) we present the positions of the U 4d$_{3/2}$ and U 4d$_{5/2}$ peaks, for which $\Delta_{SO}$ = 42.4 eV, having small intensities (see the figure caption) in relation to those in the VB region. The $\Delta_{SO}$ value is almost like that in α-U (42.1 eV) [53].

One of the most important test of band structure calculations is just a comparison of simulated and measured photoemission spectra in the energy region of VB. We note that the spectra were calculated using the results of both Wien2k and FPLO codes. As we can see in Fig. 7, a general shape of all simulated spectra is similar to one another differing only in the main peak widths (note the gray strip in this figure). Their full width at half maximum (FWHM) varies between 0.8 and 1.4 eV. Thus, the inset to Fig. 7 presents results of two codes and differences between them are clearly seen in the region 1.0 – 5.5 eV. As also shown in Fig. 7, all the calculated VB spectra, despite employing different approaches discussed above, agree in a crude way with the experimental spectrum, both in the position and energy ranges and also in the general shape of the spectral bands. However, they differ in a distinct tail expanded in the above region. For the experimental spectrum, no correction including background subtraction was made. In general, this measured spectrum is principally characterised by almost symmetric sharp main peak appeared nearly at $E_F$ (coming from predominantly the U 5f electrons) which, with increasing BE above 1 eV, transforms to a strongly non-symmetric wide hump, centred at about 2 eV BE, being not visible in any of our simulated spectra. Certainly, this hump is caused by multiplet effects connected with partly localized nature of the U 5f electrons, which could not be reproduced in calculated DOS by treating these electrons as itinerant. At about 5.0 - 5.5 eV BE, the experimental curve has a slight minimum (corresponding to a pseudogap found in the calculated DOS at the same BE). Then, it goes trough a broad peak, centred at about 8 eV, which is reflected in all our calculated DOSs in the form of two *hills*, as shown in Fig 7. This band, in addition, may be obscured by a weak oxygen line (O 2s) at around 6 eV BE. It is worth to note that here, there is good agreement of our overall spectral DOS curve with those published previously [23-25]. The latter spectra were obtained on polycrystalline samples using the excitation energies of 1254 eV (Mg Kα) and 40 eV (HeII). Although some small difference is reported between them, as e.g. in a width (FWHM) of their XPS spectrum or distances of the peak maximum from $E_F$, $\Delta E_i$, their shapes are very similar to one another.

As a rule, for many uranium compounds in the valence BE region studied here, the ionisation cross-section of U 5f- electrons at energy of Al or Mg Kα radiation is much larger than those of other valence dsp electrons in a given compound, which can be learned from cross-sections for different atoms and electron states tabulated, depending on energy of radiation, in [46]. In UGe$_2$ this effect should influence also the results presented in Fig. 8, obtained for the GGA+$U$** (Wien2k) approach, yielding the narrowest FWHM. Due to these differences in cross-sections the pronounced contributions of the Ge atom, clearly seen in Fig. 1, and coming from the 4sp states are diminished with respect to the U 5f states contribution. Thus, in Fig. 8 we have marked the contributions from the uranium and germanium atoms to the total XPS spectra. This represents a region of hybridized states originated from both U and Ge atoms with somewhat higher intensity coming from Ge electrons, as one can deduce from Fig. 8.

Very similar results were also reported by Yaresko et al. [15], who compared the results of their LSDA+$U$ ($U$ = 0, 0.5, 2.0, and 4.0 eV) approach with published photoemission data of [23]. They could reproduce the latter on-off-resonance XPS spectrum solely taking $U$ amounting to about 2 eV. This result is considerably different from $U$ of 0.7 eV given by Shick and Pickett [20] in order to explain the experimental value of the ordered magnetic moment in UGe$_2$ (see Fig. 4).

The comparison of distances $\Delta E_i$, being read off from Fig. 4 of the paper [15] and ours (i = 1 and 2, respectively), is given in Fig. 9. The hatched area in this figure draws only the difference between experimentally found $\Delta E_i^{exp}$ with indices i = 1 and 2, representing, however, possible large inaccuracy in their determination. These distances concern the location of the maximum of experimental XPS peaks for the values: 1) 0.42 eV (LSDA+*U*) and 2) 0.25 eV (GGA+*U*) BE found in [15] and in this paper, respectively. The discrepancy between these two values is probably caused (apart from using different functionals) by different kind of experimental curves used. In contrast to Yaresko et al. [15], we present in Fig. 7 the direct spectral curve obtained in the XPS process. The above comparison allows one to conclude that we have obtained similar results to those given by the above authors for *U* = 1.5 eV. However, our curve for $U \geq 1.5$ eV shows a tendency to saturation, contrary to the curve based on the results of [15], $\Delta E_1 \sim U^n$, which continuously rises even with $n \sim 2$.

In agreement with our observation described above, it is interesting to note that also the spectra of high-resolution UPS [26] measured at various temperatures exhibited a lack of any temperature variation just within the broad feature located between 1.5 and 3.5 eV BE, which led the authors of [26] to the conclusion about the existence in UGe$_2$ of "incoherent part" of spectrum originated from the localised U 5f electrons.

Finally, Fig. 10 shows the U 4f core-level spectrum of UGe$_2$ and its deconvolution into components, where background was subtracted by the Tougaard method [54]. The spectrum shows a 10.9 eV SO splitting to the 4f$_{5/2}$ and 4f$_{7/2}$ components. Each of them consists of a dominant asymmetric main line and as many as its three broad satellites, denoted as sats. I, II and III. They are located in the spectrum with increasing BE's. The numerical values of the peak positions are also collected in Fig. 10. Usually, for a number of binary and ternary intermetallic uranium compounds two peaks: main line accompanied by sat. III (so-called 7-eV satellite) are observed. Here, the positions of sats. III are separated by 7.35 and 7.85 eV BE from the 4f$_{7/2}$ and 4f$_{5/2}$ lines, respectively. As expected, the former satellite has the smallest intensity among the three ones and is symmetric but the latter one is strongly influenced by the Ge LMM line. Moreover, in this case, we have observed also two additional satellites, i.e. at about 1-eV (sat. I) and 3-eV (sat. II) in higher BE side with respect to their main lines. In such a situation, the full intensity becomes distributed to all three satellites. In turn, the origin of the sat. II, here being rather symmetric, has been not clarified as yet. Similar satellites have also been reported in previous works on UGe$_2$ [55] as well as on several other binary and ternary uranium compounds. For example, in the pseudobinary solid solutions URh$_{1-x}$Pd$_x$ the authors of [56] interpreted the presence of such a satellite as a result of a contribution, except for the two final states 5f$^2$ (sat. III) and 5f$^4$ (main line), also of the 5f$^3$ (sat. II) final state, like that existing in the 4f-core spectrum of UO$_2$ [57]. So that the latter system can be also attributed to the initial 5f$^2$ configuration of the U$^{+4}$ state. A similar situation was also met, e.g. in such a heavy-fermion ternary system as U*M*$_2$Al$_3$ (*M*= Ni and Pd) [58]. If it were a real case, we could certainly deal here with some kind of a mixed-valence character of uranium atoms in UGe$_2$. Another possibility is e.g. a *dual* character of 5f electrons, i.e. the simultaneous presence of itinerant, hybridized 5f6d states as well as almost localized 5f$^n$ states (n~2). Typical positions of main lines, 4f$_{7/2}$ and 4f$_{5/2}$, for compounds with the U$^{+3}$ ion are 388.5 and 377.8 eV BE, and with the U$^{+4}$ ion are equal to 390.9 and 380.0 eV BE, respectively. Nevertheless, the contribution of the 3-eV ($5f^3$) satellite is probably enhanced by a slight presence of uranium oxides [55]. Furthermore, as seen from the discussed figure, the shape of sat. I, called hereafter as the 1-eV satellite is asymmetric and possessing the highest intensity among three of them. It should be noted that this sat. I has not been observed in the former XPS experiments in UGe$_2$ [55]. Interestingly that solely such a satellite exists, except for the main line, in the 4f-core spectrum of spin-fluctuators, like URu(Al;Ga) [59]. In UGe$_2$, sat. I is located very closely to a given main line, i.e. at about 0.9 eV BE.

The main line can be fitted according to the Doniach-Šunjić theory [60], where the degree of asymmetry of the line is described by *the singularity index* α, which is a function of angular-dependent partial screening charges. The origin of so many satellites in the 4f-core spectrum of a given compound has not been well recognised so far, but certainly it reflects the various valence states of the U atom in a number of uranium compounds, just owing to a nature of their 5f electrons.

## 4. Conclusions

We have computed the fully relativistic, spin- and orbital-polarised electronic structure of UGe$_2$ using a variety of computational methods in order to compare the obtained results with the experimental values of the ordered magnetic moment of uranium as well as with the measured XPS spectrum (Al Kα radiation). Contrary to the previous works, reporting results of applying LSDA+*U* approach in UGe$_2$ [15,20], we obtained a very good agreement between calculated and XPS data taking a realistic (higher) value of the *U* parameter within the GGA+*U* scheme. On the other hand, the GGA+*U* method failed in the prediction of the easy magnetization axis in UGe$_2$. Therefore, we employed *ab-initio* approaches containing orbital polarization correction, as LSDA+OP and GGA+OP ones, within the FPLO and Wien2k codes, respectively. As a result, we yielded the proper ground state in both the magnetic moment value and easy magnetization axis, without introducing any external parameters as *U* and *J*. Thus, we consider these results as the most important achievement of our study. Moreover, we fully determined the Fermi surface in the FM ground state based on the LSDA+OP results. Due to this, an existence of the FS nesting vector was proved, being similar to that reported in [20] but obtained by a quite different approach.

Finally, the nature of the U 5f electrons in UGe$_2$ has been deduced from our XPS spectra. The experimental valence-band XPS spectrum, apart from the presence of the sharp peak at the Fermi level originating from the itinerant U 5f electrons, shows a pronounced hump at higher B.E. The hump is the most likely caused by multiplet effects connected with partly localized nature of the 5f electrons. The complex XPS spectrum of the 4f electrons with as many as three satellites points also to both itinerant and localized nature of the U 5f electrons. Thus, the existence of two subsets of 5f electrons in UGe$_2$, i.e. hybridized itinerant and more localized ones, reveals a dual character of these electrons. This causes close similarity of UGe$_2$ behaviour to that of UPd$_2$Al$_3$, for which such a situation was undoubtedly proved experimentally in [61]. Therefore, it seems that theoretically only such computations as those based on the self-interaction–corrected local spin density approximation (SIC-LSDA) (see e.g. [62]) might more adequately tread the problem of 5f-electron duality in this and other uranium compounds, but now it is out of our scope.

**Acknowledgements**

This work in part was supported by the Ministry of Science and Higher Education within the research projects No. N N202 1349 33 and in the frame of the National Network "Strongly correlated materials: preparation, fundamental research and applications". We are thankful for technical assistance by U. Nitzsche with IFW-Dresden computers as well as by the Computing Centre of Institute of Low Temperature and Structure Research PAS in Wrocław.

**References**

[1] S.S. Saxena, P. Agarwal, K. Ahilan, F.M. Grosche, R.K.W. Haselwimmer, M.J. Steiner, E. Pugh, I.R. Walker, S.R. Julian, P. Monthoux, G.G. Lonzarich, A. Huxley, I. Sheikin, D. Braithwaite, J. Flouquet, Nature 406 (2000) 587.
[2] A. Huxley, I. Sheikin, E. Ressouche, N. Kernavanois, D. Braithwaite, R. Calemczuk, J. Flouquet, Phys. Rev. B 63 (2001) 144519.
[3] K. Machida, T. Ohmi, Phys. Rev. Lett. 86 (2001) 850.
[4] G. Oomi, T. Kagayama, Y. Ōnuki, J. Alloys Compd. 271-273 (1998) 482.
[5] R. Troć, J. Magnetics (Korea) 9 (2004) 89.
[6] A. Pikul, R. Troć, in: Booklet of Abstracts of 39 Journées des Actinides (39JdA), La Grande Motte 28-31.03.2009, p. 170.
[7] F. Hardy, C. Meingast, V. Taufour, J. Louquet, H.v. Lõhneysen, R.A. Fisher, N.E. Phillips, A. Huxley, J.C. Lashley, Phys. Rev. B 80 (2009) 174521.
[8] Y. Ushida, H. Nakane, T. Nishioka, G. Motoyama, S. Nakamura, N.K. Sato., Physica C 388-389 (2003) 525.
[9] S. Watanabe, K. Miyake, J. Phys. Soc. Jpn. 71 (2002) 2489.
[10] N. Tateiwa, K. Hanazono, T.C. Kobayashi, K. Amaya, T. Inoue, K. Kindo, Y. Koike, N. Metoki, Y. Haga, R. Settai, Y. Ōnuki, J. Phys. Soc. Jpn. 70 (2001) 2876.
[11] K. Kuwahara, T. Sakai, M. Kohgi, Y. Haga, Y. Ōnuki, J. Magn. Magn. Mater. 310 (2007) 362.
[12] R. Troć, Acta. Phys. Polon. B 34 (2003) 407; R. Troć, J. Alloys Compd. 423 (2006) 21.
[13] V.H. Tran, S. Paschen, R. Troć, M. Baenitz, F. Steglich, Phys. Rev. B 69 (2004) 195314.
[14] S. Sakarya, P.C.M. Gubbens, A.Yaouanc, P. Dalmas de Réotier, D. Andreica, A. Amato, U. Zimmermann, N. van Dijk, E. Brück, Y. Hang, and T. Gortenulder, Phys. Rev. B 81 (2010) 02429.

[15] A.N. Yaresko, P. Dalmas de Réotier, A. Yaouanc, N. Kernavanois, J.-P. Sanchez, A.A. Menovsky, V.N. Antonov, J. Phys.: Condens. Matter 17 (2005) 2443.

[16] K. Miyake and Y. Kuramoto, J. Magn. Magn. Mater. 90-91 (1990) 438.

[17] K.G. Sandeman, G.G. Lonzarich, A.J. Schofield, Phys. Rev. Lett. 90 (2003) 167005.

[18] N. Aso, G. Motoyama, Y. Uwatoko, S. Ban, S. Nakamura, T. Nishioka, Y. Homma, Y. Shiokawa, K. Hirota, N.K. Sato, Phys. Rev. B 73 (2006) 054512.

[19] A. Szajek, Cryst. Res. Technol. 36 (2001) 1105.

[20] A.B. Shick, W.E. Pickett, Phys. Rev. Lett. 86 (2001) 300.

[21] A.B. Shick, V. Janiš, V. Drchal, W.E. Pickett, Phys. Rev. B 70 (2004) 134506.

[22] H. Yamagami, J. Phys.: Condens. Matter 15 (2003) S2271.

[23] S. Suzuki, S. Sato, T. Ejima, K. Murata, Y. Kudo, T. Takahashi, T. Komatsubara, N. Sato, M. Kasaya, T. Suzuki, T. Kasuya, S. Suga, H. Matsubara, Y. Saito, A. Kimura, K. Soda, Y. Ōnuki, T. Mori, A. Kakizaki, T. Ishii, Jpn. J. Appl. Phys. Series 8 (1993) 59.

[24] T. Ishii, Physica B 186-188 (1993) 21.

[25] K. Soda, T. Mori, Y. Ōnuki, T. Komatsubara, S. Suga, A. Kakizaki, T. Ishii, J. Phys. Soc. Jpn. 60 (1991) 3059.

[26] T. Ito, H. Kumigashira, S. Souma, T. Takahashi, Y. Haga, Y. Ōnuki, J. Phys. Soc. Jpn. (Suppl.) 71 (2002) 261.

[27] M. Biasini, Troć R., Phys. Rev. B 68 (2003) 245118.

[28] Y. Ōnuki, S.W. Yun, I. Ukon, I. Umehara, K. Satoh, I. Sakamoto, M. Hunt, P. Meeson, P.-A. Probst, M. Springford, J. Phys. Soc. Jpn. 60 (1991) 2127.

[29] K. Satoh, S.W. Yun, I. Ukon, I. Umehara, Y. Ōnuki, H. Aoki, S. Uji, T. Shimizu, I. Sakamoto, M. Hunt, P. Meeson, P.-A. Probst, M. Springford, J. Magn. Magn. Mater. 104-107 (1992) 39.

[30] K. Satoh, S.W. Yun, I. Umehara, Y. Ōnuki, S. Uji, T. Shimizu, H. Aoki, J. Phys. Soc. Jpn. 61 (1992) 1827.

[31] S.W. Yun, K. Satoh, Y. Fujimaki, I. Umehara, Y. Ōnuki, S. Takayanagi, H. Aoki, S. Uji, T. Shimizu, Physica B 186-188 (1993) 129.

[32] N. Tateiwa, T.C. Kobayashi, K. Hanazono, K. Amaya, Y. Haga, R. Settai, Y. Ōnuki, J. Phys.: Condens. Matter 13 (2001) L17.

[33] P. Boulet, A. Daoudi, M. Potel, H. Noël, G.M. Gross, G. André, F. Bourée, J. Alloys. Compd. 247 (1997) 104.

[34] P. Hohenberg, W. Kohn, Phys. Rev. 136 (1964) B864.

[35] K. Koepernik, H. Eschrig, Phys. Rev. B 59 (1999) 1743 (FPLO code versions 5.10-20 and 7.00-28; http://www.FPLO.de).

[36] P. Blaha, K. Schwarz, G.K.H. Madsen, D. Kvasnicka, J. Luitz, Wien2k, An Augmented Plane Wave Plus Local Orbitals Program for Calculating Crystal Properties, Schwarz K. (ed.), Techn. Universität Wien, Austria, 2001.

[37] J.P. Perdew, Y. Wang, Phys. Rev. B 45 (1992) 13244.

[38] A.H. MacDonald, W.E. Pickett, D.D. Koelling, J. Phys. C: Solid State Phys. 13 (1980) 2675.

[39] J.P. Perdew, K. Burke, M. Ernzerhof, Phys. Rev. Lett. 77 (1996) 3865.

[40] Z. Wu, R.E. Cohen, Phys. Rev. B 73 (2006) 235116; F. Tran, R. Laskowski, P. Blaha, K. Schwarz, Phys. Rev. B **75** (2007) 115131.

[41] (a) V.I. Anisimov, I.V. Solovyev, M.A. Korotin, M.T. Czyżyk, G.A. Sawatzky, Phys. Rev. B 48 (1993) 16929; (b) A.I. Liechtenstein, V.I. Anisimov, J. Zaanen, Phys. Rev. B 52 (1995) R5467; (c) http://www.wien2k.at/reg_user/textbooks/double_counting.ps; P. Novak, F. Boucher, P. Gressier, P. Blaha, K. Schwarz, Phys. Rev. B 63 (2001) 235114.

[42] M.S.S. Brooks, Physica B&C 130B (1985) 6.

[43] O. Eriksson, B. Johansson, M.S.S. Brooks, J. Phys.: Condens. Matter 1 (1989) 4005; C. Neise, OP implementation in the FPLO5.10-20 code, diploma thesis: Orbital Polarization Corrections in Relativistic Density Functional Theory, Technische Universität, Dresden, 2007; http://www.ifw-dresden.de/institutes/itf/diploma-and-phd-theses-at-the-itf

[44] O. Eriksson, M.S.S. Brooks, B. Johansson, Phys. Rev. B 41 (1990) 7311.

[45] P.E. Blöchl, O. Jepsen, O.K. Andersen, Phys. Rev. B 49 (1994) 16223.

[46] J.J. Yeh, I. Lindau, At. Data Nucl. Data Tables 32 (1985) 1.

[47] P. Marksteiner, P. Weinberger, R.C. Albers, A.M. Boring, G. Schadler, Phys. Rev. B 34 (1986) 6730.

[48] N. Kernavanois, B. Grenier, A. Huxley, E. Ressouche, J.P. Sanchez, J. Flouquet, Phys. Rev. B 64 (2001) 174509.

[49] J. Yu, R. Devanathan, W,J. Wener, J. Phys.: Condens. Matter 21 (2009) 435401.

[50] V.N. Antonov, B.N. Harmon, O.V. Andryuschenko, L.V. Bekenev, A.N. Yaresko, Fizika Nizkikh Temp. 30 (2004) 411.

[51] J.S. Kang, S.C. Wi, J.H. Kim, K.A. McEwen, C.G. Olson, J.H. Shim, B.I. Min, J. Phys.: Condens. Matter 16 (2004) 3257.

[52] R. Settai, M. Nakashima, S. Araki, Y. Haga, T.C. Kobayashi, N. Tateiwa, H. Yamagami, Y. Ōnuki, J. Phys.:Condens. Matter 14 (2002) L29.

[53] S. Hufners, in: Photoemission spectroscopy, Berlin, Springer, 1994, p. 453.

[54] S. Tougaard, J. Electron Spectrosc. Relat. Phenom. 52 (1990) 243.

[55] T. Ejima, S. Sato, S. Suzuki, Y. Saito, S. Fujimori, N. Sato, M. Kasaya, T. Komatsubara, T. Kasuya, Y. Ōnuki, T. Ishii, Phys. Rev. B 53 (1996) 1806, and references therein.

[56] S. Fujimori, Y. Saito, N. Sato, T. Komatsubara, S. Suzuki, S. Sato, T. Ishii, J. Phys. Soc. Jpn. 67 (1998) 4164.

[57] Y. Baer, J. Schoenes, Solid State Commun. 33 (1980) 885.

[58] S. Fujimori, Y. Saito, N. Sato, T. Komadsubara, S. Suzuki, S. Sato, T. Ishii, Solid State Commun. 105 (1998) 185.

[59] M. Samsel-Czekała, E. Talik, R. Troć, Phys. Rev. B 78 (2008) 245120; M. Samsel-Czekała, E. Talik, R. Troć, J. Stępień-Damm, Phys. Rev. B 77 (2008) 155113..

[60] S. Doniach, M. Šunjić, J. Phys. C: Solid State Phys. 3 (1970) 285.

[61] N.K. Sato, N. Aso, K. Miyake, R. Shiina, P. Thalmeier, G. Varelogiannis, G. Geibel, F. Steglich, P. Fulde, T. Komatsubara, Nature 410 (2001) 340.

[62] L. Petit, A. Svane, W.M. Temmerman, Z. Szotek, R. Tyer, Europhys. Lett. 62 (2003) 391.

**Tables**

**Table 1.** Calculated values of total ordered magnetic moment, $\mu_{tot}$, and the ratio of its orbital and spin contributions, $\mu_l/\mu_s$, contributions, DOSs at $E_F$, Sommerfeld coefficient, $\gamma_b$, and the magnetocrystalline anisotropy energy, MAE, relative to easy axis: $\mu$ [$\mu_B$/U atom]; DOS(E=$E_F$) [states/(eV spin f.u.)]; $\gamma_b$ [mJ K$^{-2}$ mol$^{-1}$]; MAE (mRy); arrows ↑, ↓ mean spin up and down, respectively. The results were obtained for FPLO (a) and Wien2k (b) codes within the LSDA+OP and GGA+$U$ schemes, respectively.

| axis | $\mu_{tot}$ | $\mu_l/\mu_s$ | DOS($E_F$) | $\gamma_b$ | MAE |
|---|---|---|---|---|---|
| (a) | | FPLO | | LSDA+OP | |
| [001] | 1.53 | -1.85 | ↑3.95 ↓1.34 | 12.46 | 2.42 |
| [010] | 1.49 | -1.83 | ↑4.25 ↓1.29 | 13.05 | 2.32 |
| [100] | 1.61 | -1.93 | ↑2.79 ↓1.23 | 9.47 | 0.00 |
| (b) | | Wien2k | | GGA+OP | |
| [001] | 1.50 | -1.90 | ↑4.35 ↓0.89 | 12.35 | 41.11 |
| [010] | 1.46 | -1.87 | ↑5.68 ↓0.91 | 15.54 | 40.62 |
| [100] | 1.56 | -1.97 | ↑7.95 ↓0.88 | 20.80 | 0.00 |

**Table 2.** Same as in Table1, but for results obtained for the Wien2k code within the GGA+$U$ approach for different values of $U$ and $J$ parameters (note a change in the easy magnetisation direction).

| | $U$= 0.7 eV; $J$= 0.44 eV |
|---|---|

| axis | $\mu_{tot}$ | $\mu_l/\mu_s$ | DOS($E_F$) | $\gamma_b$ | MAE |
|---|---|---|---|---|---|
| [001] | 0.91 | -1.54 | ↑5.49 ↓0.82 | 14.86 | 1.52 |
| [010] | 1.03 | -1.60 | ↑4.50 ↓0.92 | 12.79 | 0.06 |
| [100] | 1.09 | -1.68 | ↑5.81 ↓0.93 | 15.89 | 0.00 |
| axis | | $U$= 2.0 eV; $J$= 0.55 eV | | | |
| [001] | 1.34 | -1.68 | ↑4.22 ↓0.75 | 11.73 | 1.68 |
| [010] | 1.20 | -1.57 | ↑3.67 ↓0.79 | 10.53 | 0.00 |
| [100] | 1.31 | -1.65 | ↑4.74 ↓0.72 | 12.87 | 3.00 |
| axis | | $U$= 3.0 eV; $J$= 0.55 eV | | | |
| [001] | 1.48 | -1.57 | ↑4.27 ↓0.80 | 11.94 | 3.36 |
| [010] | 1.22 | -1.54 | ↑3.16 ↓0.76 | 9.24 | 0.00 |
| [100] | 1.41 | -1.67 | ↑4.64 ↓0.68 | 12.54 | 10.86 |

**Figure captions**

**Fig. 1.** (Color online) Total and partial (the latter for given kind of atom and/or electron orbital) DOSs plots for UGe$_2$ obtained by FPLO in the LDA approach. For better visualisation, some lines are, additionally, numbered.

**Fig. 2.** (Colour online) Total DOSs plots for spin polarisation along the easy magnetisation axis [100] in UGe$_2$ obtained by the FPLO and Wien2k codes within the LSDA+OP (a) and GGA+$U$ (b) approaches.

**Fig. 3.** (Colour online) Same as in Fig. 2 but at the vicinity of the Fermi level (in expanded scale) and compared with the respective results of calculations with spin polarisation along a hard magnetisation axis [001].

**Fig. 4.** (Colour online) Dependence of the calculated total magnetic moments, $\mu_{tot}$, in UGe$_2$ on the Coulomb $U$ parameter taken in our GGA+$U$ approach within our Wien2k calculations (see Table 2) and in the LSDA+$U$ computations, presented in [15] and [20], marked by closed and open squares, respectively. For comparison, also LSDA, LSDA+OP and GGA+OP values are drawn in the column (see Table 1) and experimental results of [48, 33] are marked by black lines.

**Fig. 5.** (Colour online) Calculated for UGe$_2$ largest FS sheet in the 136 band, in the ferromagnetically ordered state along the [100] direction, based on the LSDA+OP approach in the FPLO code: (a) drawn in the conventional u.c. (in the reciprocal space) and (b) its section in *ac* plane with marked by dark arrow visualizing a possible nesting vector **q**= $2\pi/a$ (0.47,0,0).

**Fig .6.** (Colour online) Experimental XPS spectrum for UGe$_2$ in the range of Ge 3d line accompanied by U 6p peaks (a) and U 4d lines (b), the latter spectrum was obtained by long time exposure, due to small intensity. The inset to part (a) shows decomposition of the Ge 3d line into two peaks split due to SO.

**Fig. 7.** (Colour online) Comparison of measured and simulated XPS spectra obtained by FPLO (LDA) and Wien2k (GGA, GGA+$U$) codes. The latter for three sets of $U$ and $J$ parameters: $U^*$= 0.7 eV and $J$= 0.44 eV; $U^{**}$= 1.5 eV and $J$= 0.55 eV; $U^{***}$= 2.0 eV and $J$= 0.55 eV. One can deduce from the inset that the narrowest and broadest main XPS peaks were yielded by Wien2k code in the GGA+$U^{**}$ and GGA+$U^{***}$ schemes, respectively. The resulting difference between this two cases is marked by the hatched grey area.

**Fig. 8.** (Colour online) Simulated total XPS spectrum based on: Wien2k (GGA+$U^{**}$) method, decomposed into the contributions of U and Ge atoms. For comparison the experimental spectrum is also shown in the figure.

**Fig. 9.** (Colour online) Distances of the main U 5f-electron peak from the Fermi level, $\Delta E_i$, in the XPS spectrum of UGe$_2$ depending on the $U$ parameter taken in: (i=1) LMTO simulations of [15]; (i=2) Wien2k (GGA+$U$) approach. For comparison, also our experimental results and that of [23] are marked by black lines numbered as 2 and 1, respectively.

**Fig. 10.** (Colour online) U 4f core-level experimental spectrum of UGe$_2$ with the background subtracted and its decomposition into the main line and three satellites (I-III).

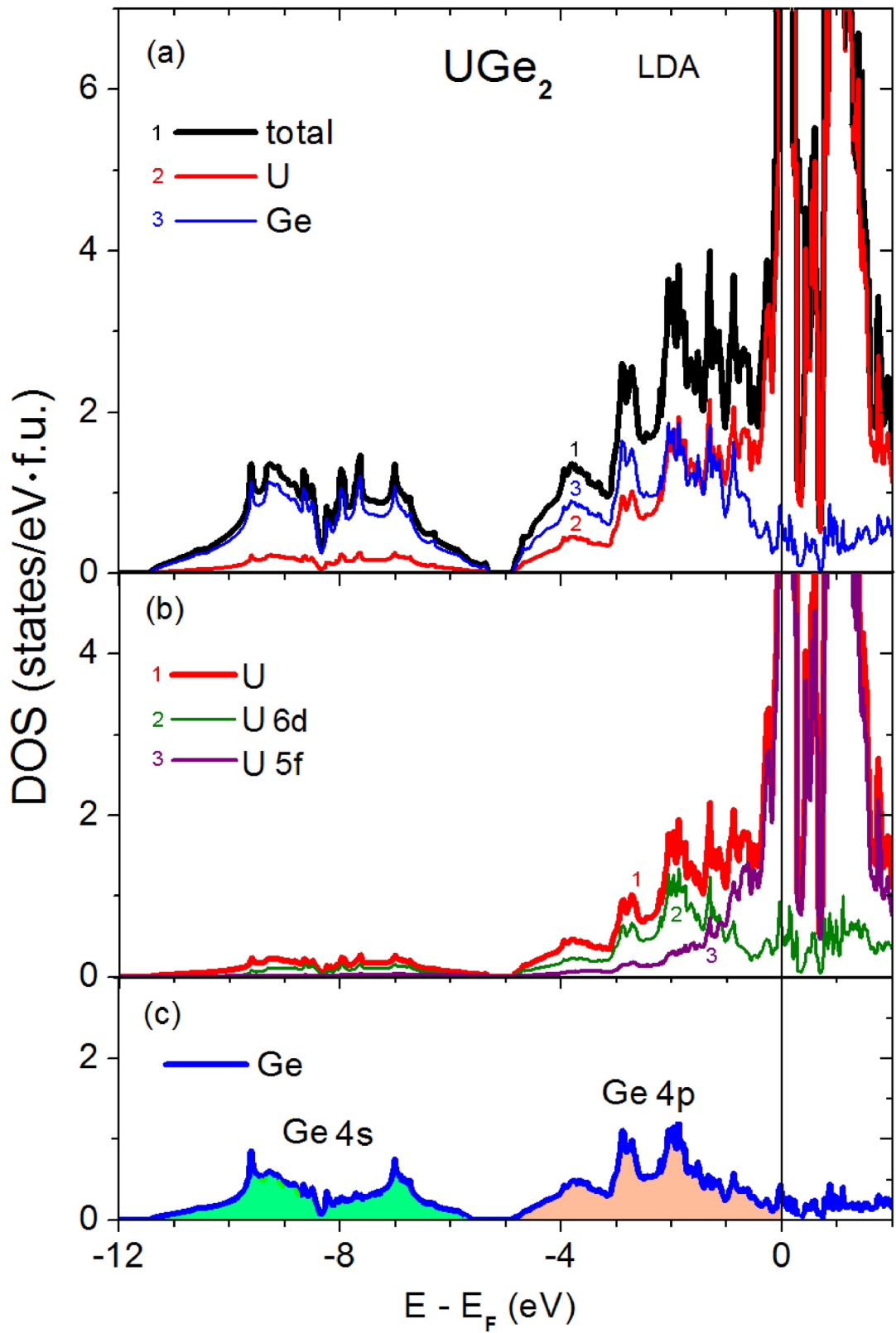

*Figure 1: Total and partial (the latter for given kind of atom and/or electron orbital) DOSs plots for UGe$_2$ obtained by FPLO in the LDA approach. For better visualisation, some lines are, additionally, numbered.*

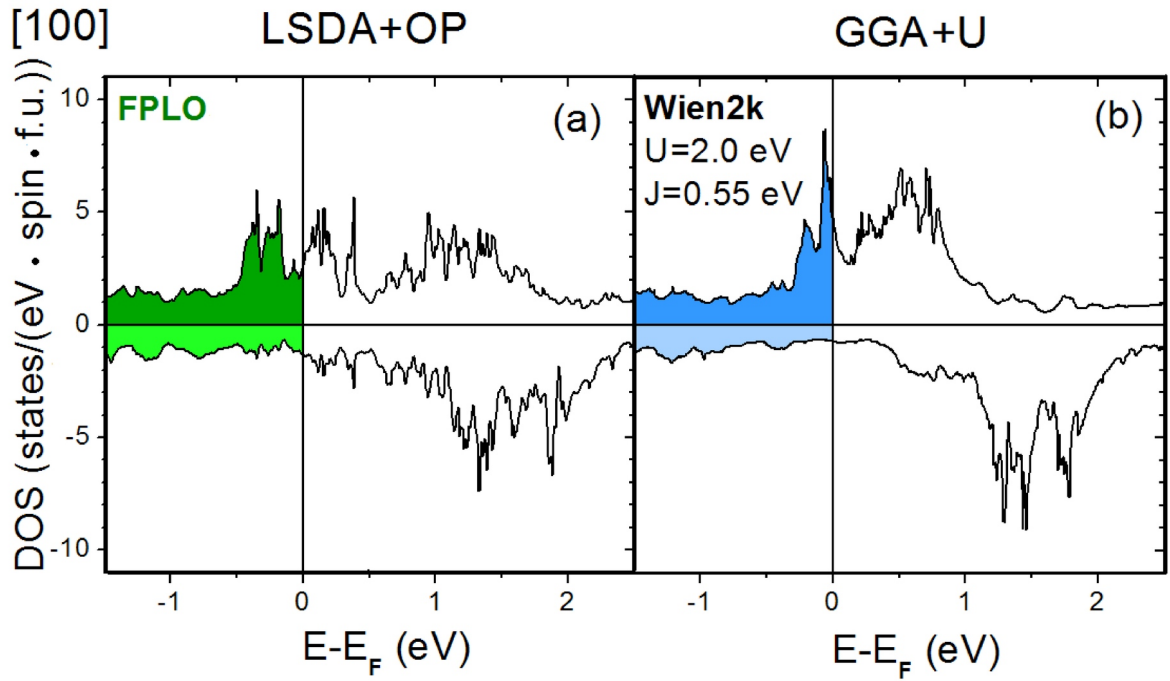

*Figure 2: Total DOSs plots for spin polarisation along the easy magnetisation axis [100] in UGe$_2$ obtained by the FPLO and Wien2k codes within the LSDA+OP (a) and GGA+U (b) approaches.*

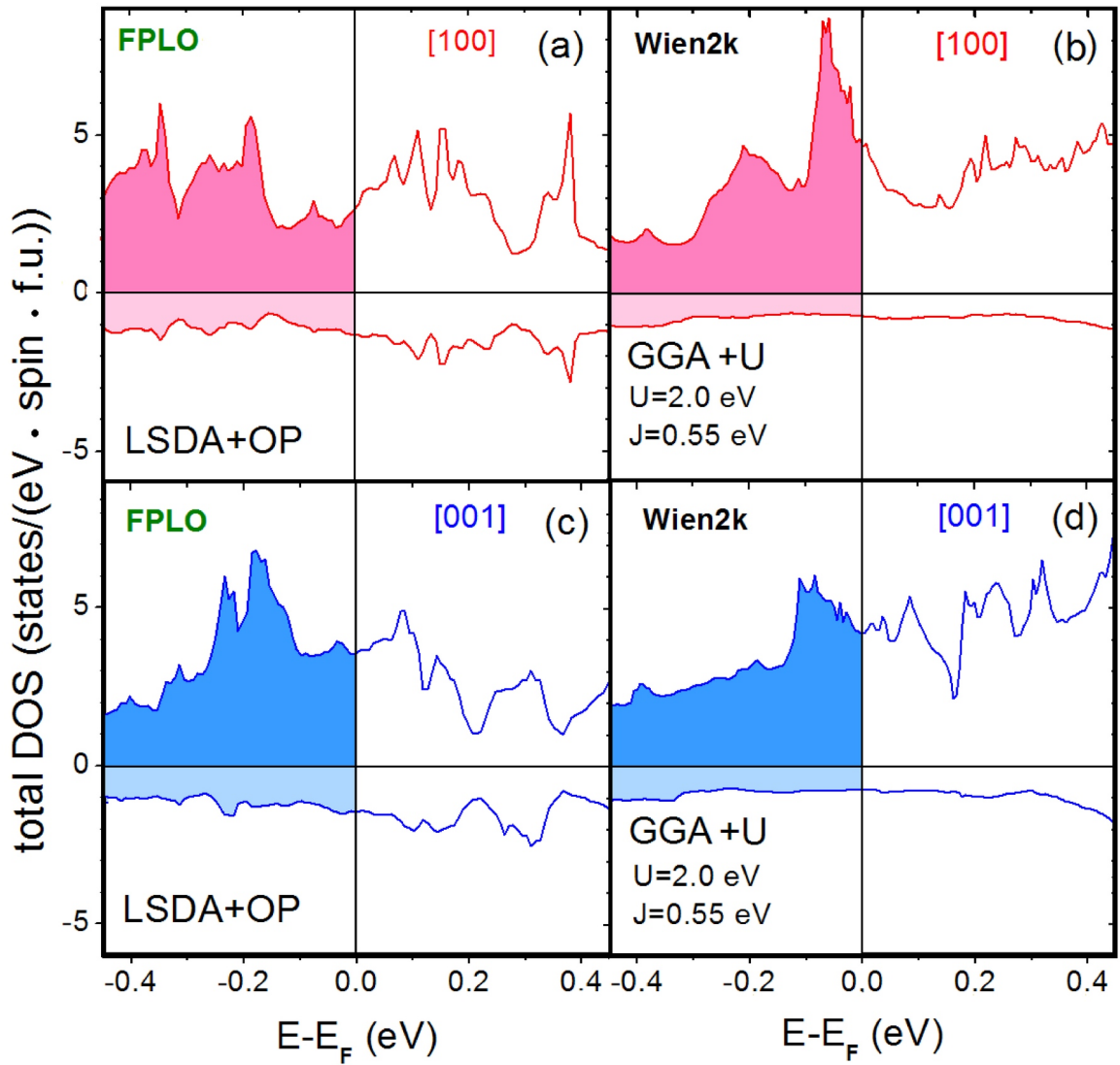

*Figure 3: Same as in Fig. 2 but at the vicinity of the Fermi level (in expanded scale) and compared with the respective results of calculations with spin polarisation along a hard magnetisation axis [001].*

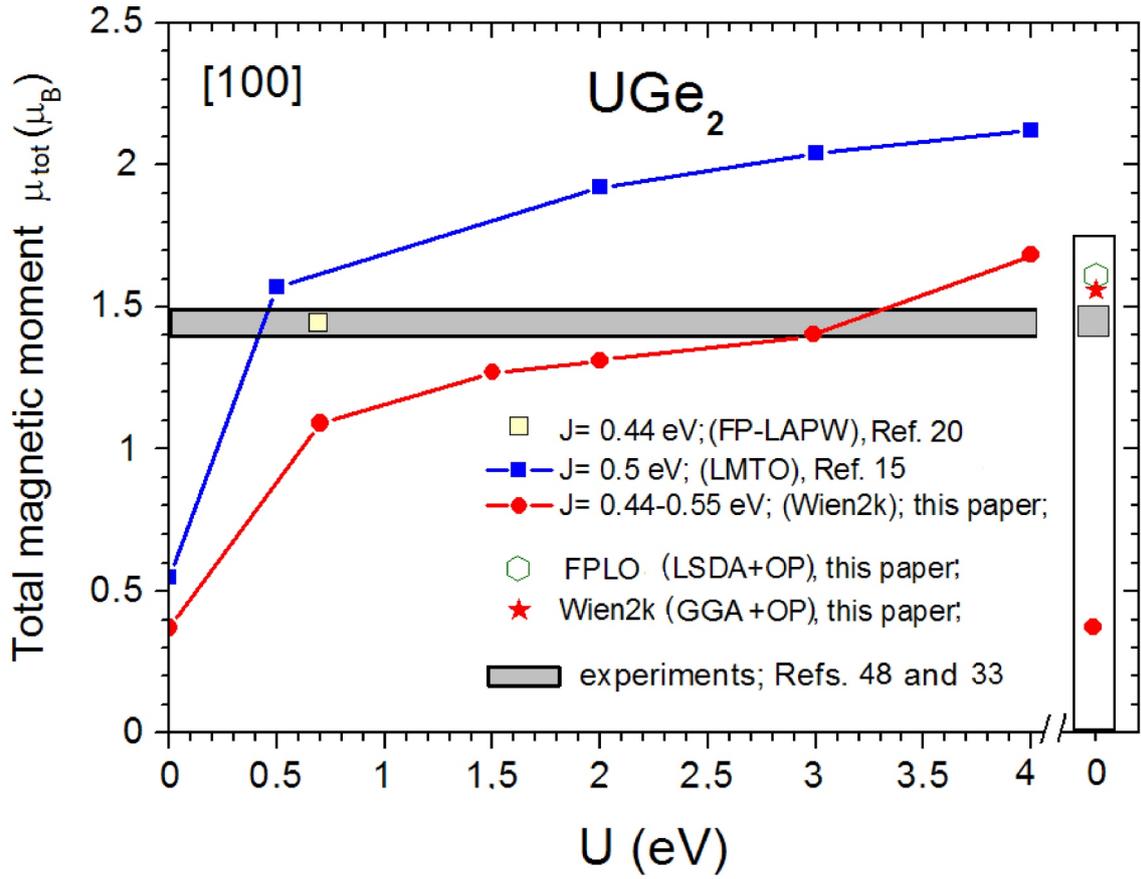

*Figure 4: Dependence of the calculated total magnetic moments, $\mu_{tot}$, in UGe$_2$ on the Coulomb U parameter taken in our GGA+U approach within our Wien2k calculations (see Table 2) and in the LSDA+U computations, presented in [15] and [20], marked by closed and open squares, respectively. For comparison, also LSDA, LSDA+OP and GGA+OP values are drawn in the column (see Table 1) and experimental results of [48, 33] are marked by black lines.*

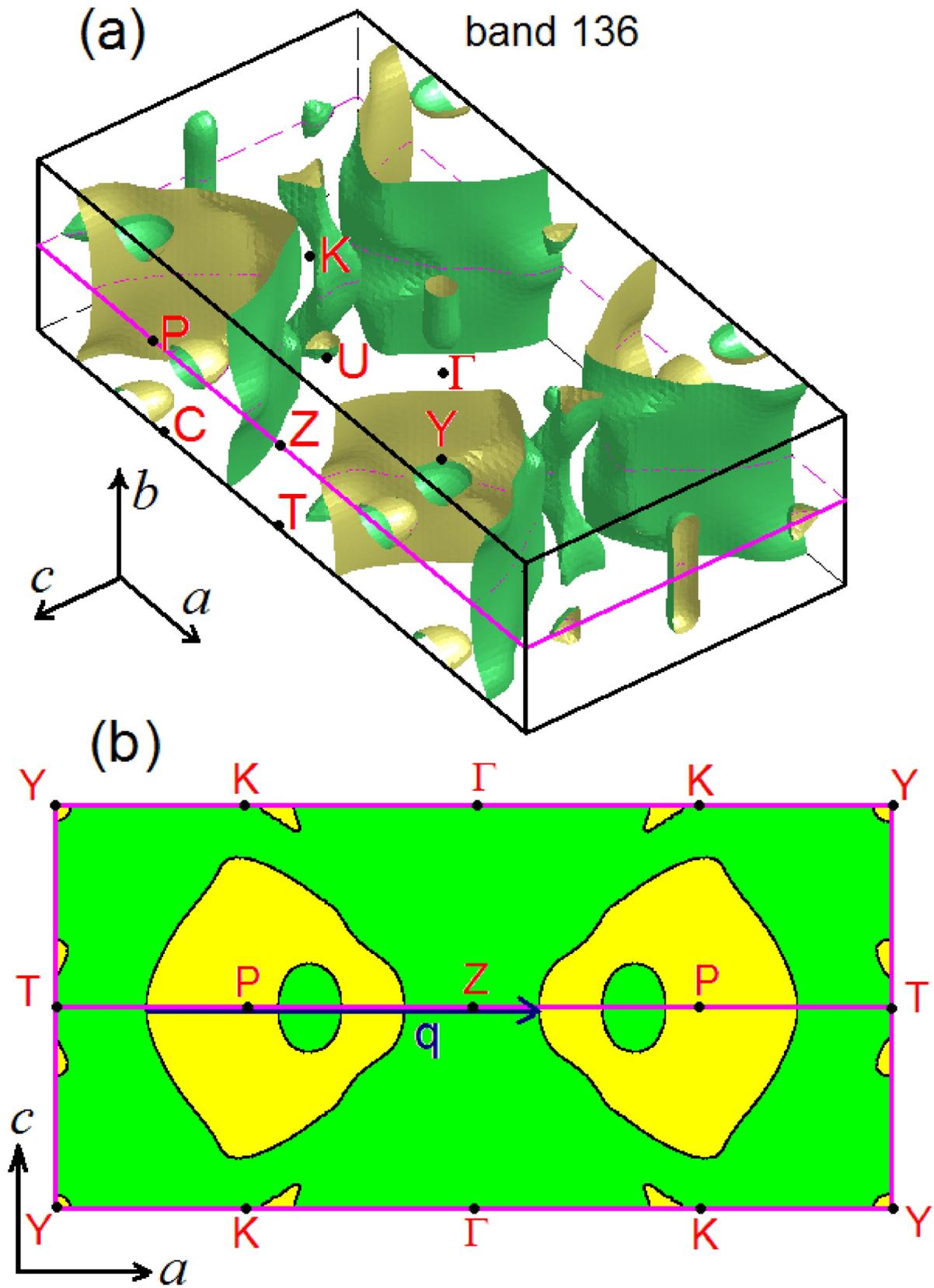

*Figure 5: Calculated for UGe$_2$ largest FS sheet in the 136 band, in the ferromagnetically ordered state along the [100] direction, based on the LSDA+OP approach in the FPLO code: (a) drawn in the conventional u.c. (in the reciprocal space) and (b) its section in ac plane with marked by dark arrow visualizing a possible nesting vector **q**= 2π/a (0.47,0,0).*

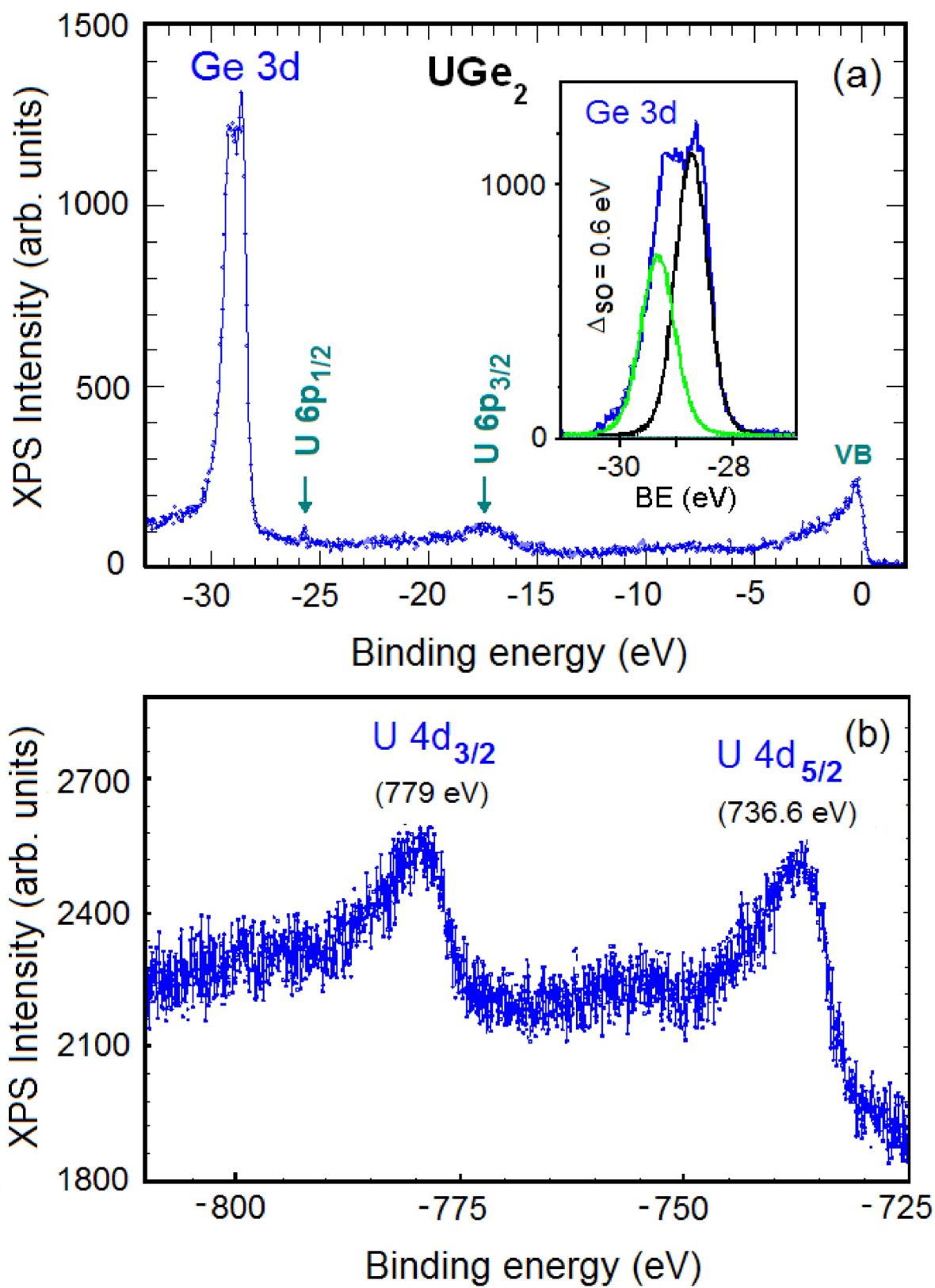

*Figure 6: Experimental XPS spectrum for UGe$_2$ in the range of Ge 3d line accompanied by U 6p peaks (a) and U 4d lines (b), the latter spectrum was obtained by long time exposure, due to small intensity. The inset to part (a) shows decomposition of the Ge 3d line into two peaks split due to SO.*

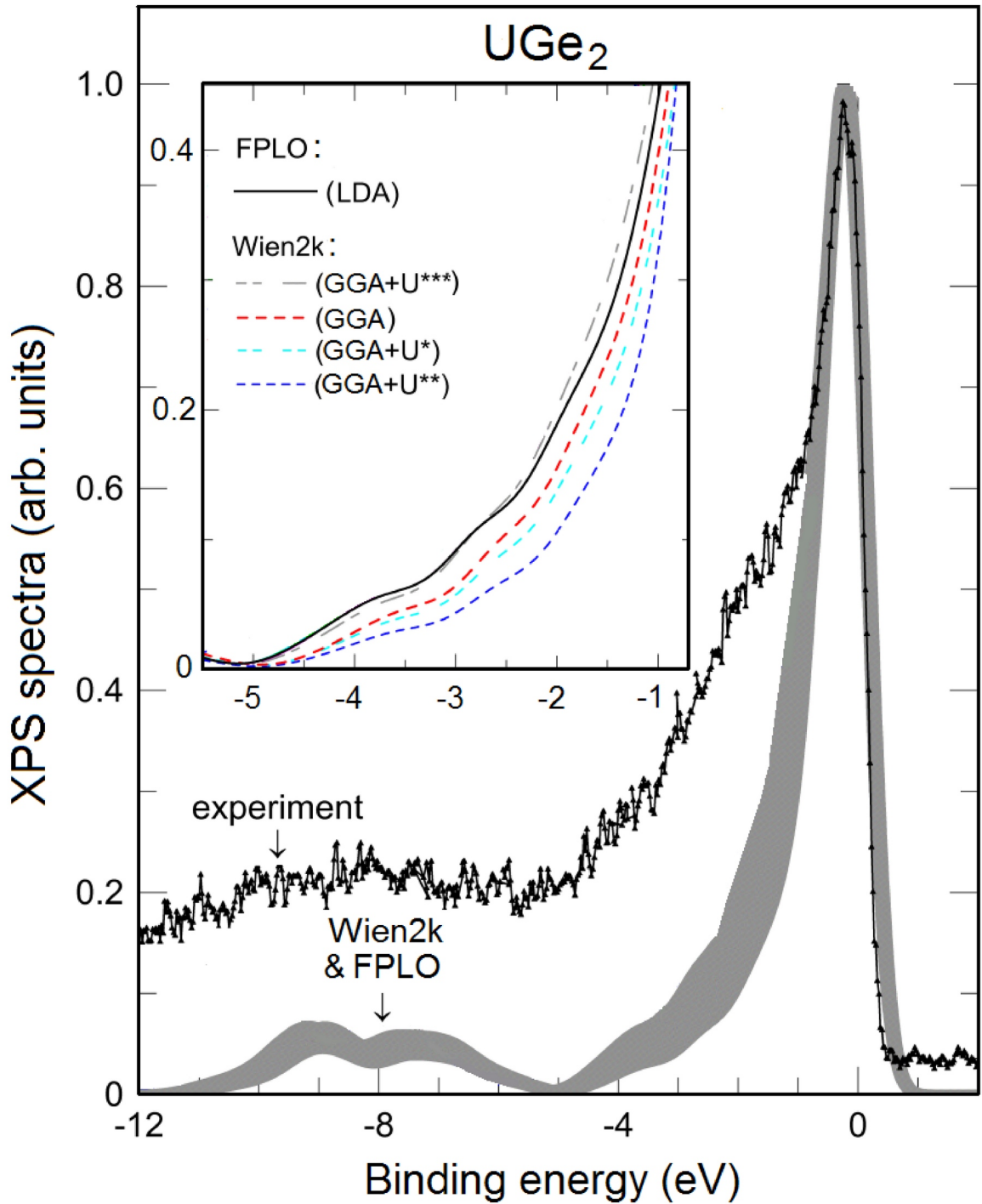

*Figure 7: Comparison of measured and simulated XPS spectra obtained by FPLO (LDA) and Wien2k (GGA, GGA+U) codes. The latter for three sets of U and J parameters: U\*= 0.7 eV and J= 0.44 eV; U\*\*= 1.5 eV and J= 0.55 eV; U\*\*\*= 2.0 eV and J= 0.55 eV. One can deduce from the inset that the narrowest and broadest main XPS peaks were yielded by Wien2k code in the GGA+U\*\* and GGA+U\*\*\* schemes, respectively. The resulting difference between this two cases is marked by the hatched grey area.*

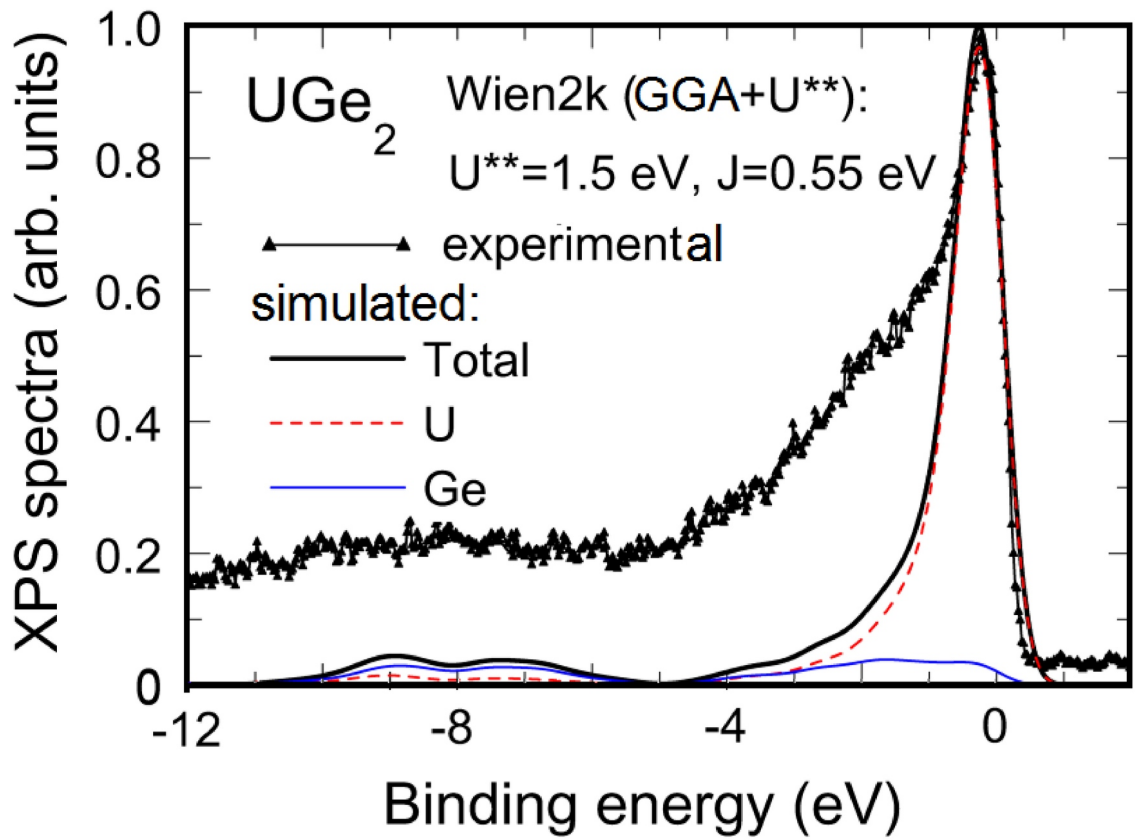

*Figure 8: Simulated total XPS spectrum based on: Wien2k (GGA+U\*\*) method, decomposed into the contributions of U and Ge atoms. For comparison the experimental spectrum is also shown in the figure.*

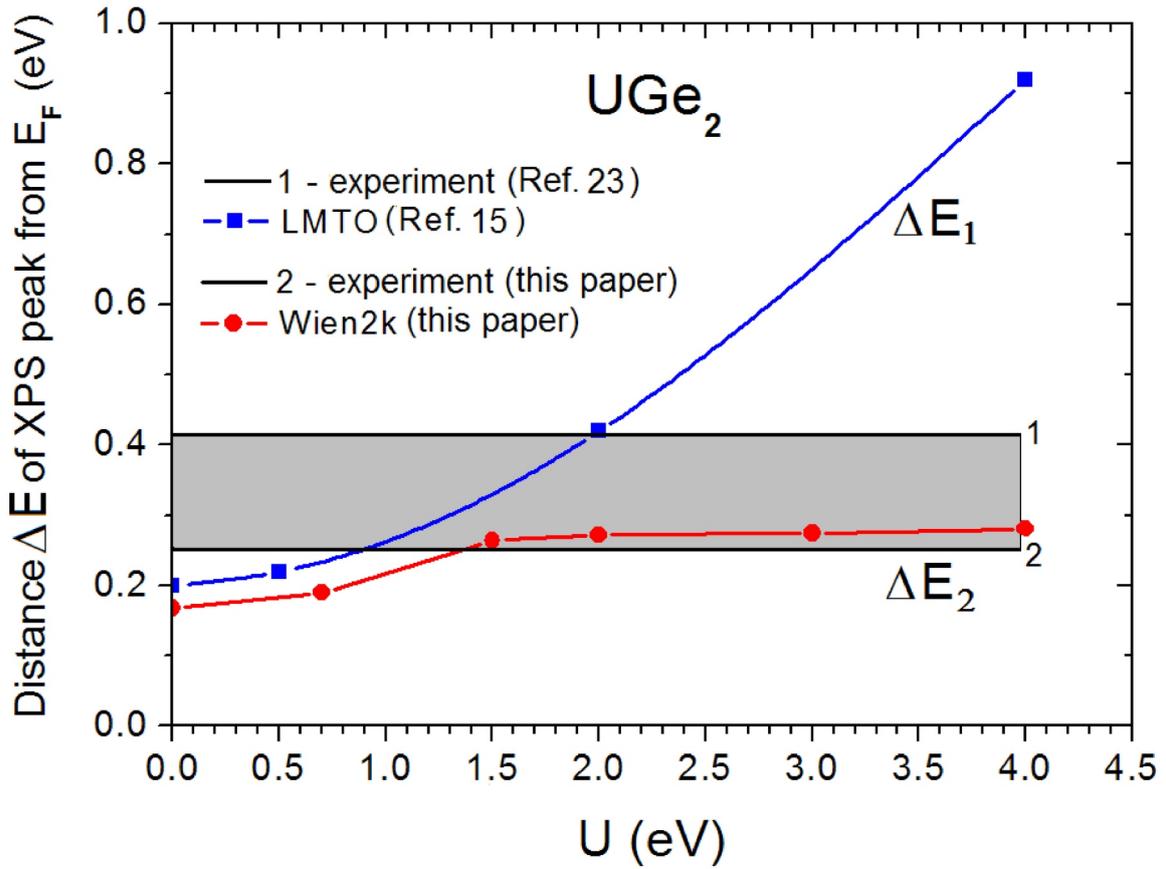

*Figure 9: Distances of the main U 5f-electron peak from the Fermi level, $\Delta E_i$, in the XPS spectrum of $UGe_2$ depending on the U parameter taken in: (i=1) LMTO simulations of [15]; (i=2) Wien2k (GGA+U) approach. For comparison, also our experimental results and that of [23] are marked by black lines numbered as 2 and 1, respectively.*

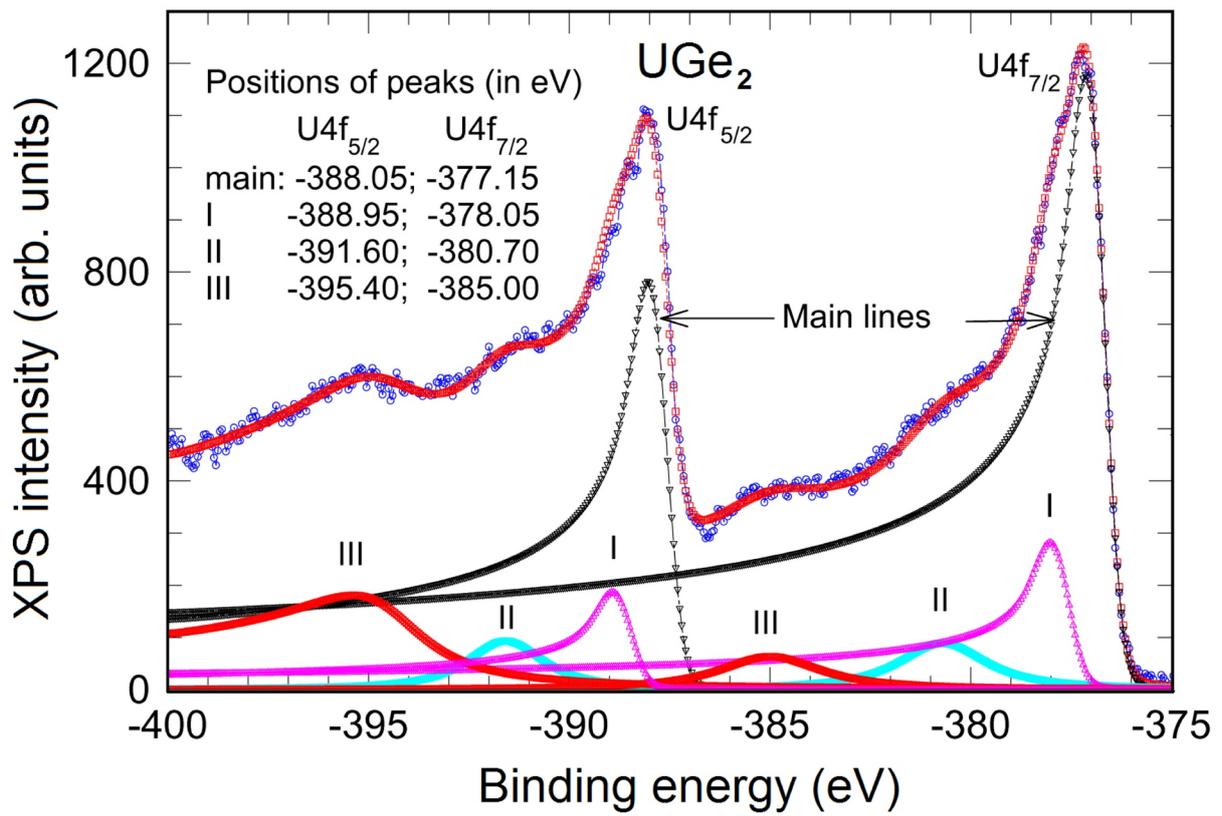

*Figure 10: U 4f core-level experimental spectrum of UGe₂ with the background subtracted and its decomposition into the main line and three satellites (I-III).*